\documentclass[times,11pt]{article}
\pdfoutput=1
\usepackage{amsmath,amssymb,amsfonts,latexsym,amsthm,enumerate,url}
\usepackage{latex8}
\usepackage{times}
\usepackage{graphicx} 
\date{}

\newtheorem{thm}{Theorem}[section]
\newtheorem{theo}[thm]{Theorem}

\newtheorem{prop}[thm]{Proposition}

\newcommand{\beq}[1]{\begin{equation}\label{#1}}
\newcommand{\enq}[0]{\end{equation}}

\newcommand{\remove}[1]{}

\begin{document}
\title {How Quantum Computers Fail:\\ Quantum Codes, Correlations in 
Physical Systems, and Noise Accumulation}

\author{ {\large Gil Kalai}\thanks{kalai@math.huji.ac.il, Work supported by a BSF grant 
and by an NSF grant.} 
\\ {\small The Hebrew University of Jerusalem,}\\ 
{\small Microsoft Research, Hertzelia,}\\ {\small and Yale University}
} 
\maketitle
\begin {center}
{\em Dedicated to the memory of Itamar Pitowsky}
\end {center}

\begin{abstract} 
The feasibility of computationally superior quantum computers 
is one of the most exciting and clear-cut scientific questions of our time.
The question touches on fundamental issues regarding probability, 
physics, and computability, as well as on exciting problems in 
experimental physics, engineering, computer science,  
and mathematics. We propose three 
related directions towards a negative answer. The first is 
a conjecture about physical realizations of quantum codes, the second  
has to do with correlations in stochastic physical systems, and the third 
proposes a model for quantum evolutions when noise accumulates.

\end{abstract}



\section {Introduction}
\label {s:in}

Quantum computers were offered by Feynman \cite {Fey} and others and 
were formally described by Deutsch \cite {Deu}. Feynman and Deutsch proposed
a bold conjecture:

\begin {itemize}
\item[]
{\bf The postulate of quantum computation:} 
{\it Computational devices based on quantum mechanics 
will be computationally superior compared to 
digital computers.} 
\end {itemize}

The idea was that since computations in 
quantum physics 
require an exponential number of steps on digital 
computers, computers based on quantum physics 
may outperform classical computers. A spectacular support for this 
idea came with Peter Shor's theorem \cite {S1} 
that asserts that quantum computers can factor integers in polynomial time. 
It is not known if there is a classical algorithm to factor $n$-digit 
integers in polynomial time, and it is widely believed that this is impossible.
Moreover, much of modern cryptography, as well as security in computer 
systems for 
finance and commerce are based on the 
assumption that factoring integers is computationally hard. 

The feasibility of computationally superior quantum computers is one
of the most fascinating and clear-cut scientific problems of our time. 
The main concern regarding quantum-computer feasibility is that 
quantum systems are noisy. Especially damaging is decoherence, 
which amounts to information leaks from a quantum system to its environment.

\begin {itemize}
\item[]
{\bf The postulate of noise:} {\em Quantum systems are inherently noisy.} 
\end {itemize}

The postulate of noise and the nature of 
decoherence are 
intimately related to questions about the nature and origins of 
probability, uncertainty, and approximations in physics.
The  concern regarding noise was put forward in
the mid-90s by Landauer \cite {lan,lan2}, Unruh \cite {unr}, and others.  
The theory of quantum error correction and fault-tolerant quantum 
computation (FTQC)
and, in particular, the {\it threshold theorem} \cite{AB2,Kit1,KLZ}, 
which asserts that 
under certain conditions FTQC is possible, 
provide strong support for the possibility of building quantum
computers. Our next Section \ref {s:bm} describes the basic 
framework for quantum computers, 
noisy quantum computers, and the threshold theorem.  
The emerging theory of quantum fault tolerance 
gives hope for building quantum computers but at the same time 
raises some concern regarding the initial rationale
of Feynman. 
As far as we know, quantum error correction and 
quantum fault tolerance (and the very highly entangled quantum 
states that enable them\footnote{For a mathematical distinction 
between the type of entangled states we encounter in nature and 
those very entangled states required for quantum fault tolerance, 
see Conjecture C in \cite {K,K:wna}}) are not experienced 
in natural quantum processes. 
Our understanding that computationally superior quantum computers
depend on realizing quantum error correction codes, which are not witnessed
in nature, weakens  the initial rationale given for quantum computers. 
It is not clear if computationally superior quantum 
computation is necessary to describe natural quantum processes, 
or, in other words, if there are {\it computational} obstruction for 
simulating natural quantum processes on digital computers.

In this paper we address two closely related questions. 
The first is  
what kind of noise models cause
quantum error correction and FTQC to fail. 
The second is what are the properties of quantum processes that 
do not exhibit quantum fault tolerance and how are 
such processes formally modeled. We discuss three related 
directions. The first deals with quantum codes, the second deals with 
correlation and entanglement of
noisy stochastic physical systems, and the third deals with modeling 
noisy quantum evolutions when noise accumulates.

Quantum error correcting codes are at the heart of the issue. 
We will discuss them in Section \ref {s:codes}.
The hope regarding FTQC is that no matter what the quantum computer 
computes or simulates, 
nearly all of the noise will be a mixture of states that are 
not codewords in the error correcting code, but which are 
correctable to states in the code.  
The concern regarding quantum codes 
is expressed by the following conjecture: 



\begin {quote}
{\bf Conjecture 1:} {\em The process for creating a quantum error correcting 
code will necessarily lead to a mixture of the 
desired codeword with undesired codewords.} 
\end {quote}

A main point we would like to make 
is that it is 
possible that there is  a systematic relation between 
the noise and the intended state of a quantum computer. 
Indeed, Conjecture 1
proposes such a systematic relation. 
Such a  
relation 
does not violate linearity of quantum mechanics, and
it is expected to occur in processes that do not exhibit fault tolerance. 
Let me give an example: suppose that we want to simulate on a 
noisy quantum computer a certain 
bosonic state. 
The standard view of noisy quantum computers asserts that 
this can be done up to some error that strongly depends on 
the computational basis. (The computational basis is a basis of the 
Hilbert space based on the individual qubits.) In contrast, we can regard
the subspace of bosonic states 
as a quantum code, and the type of noise we expect 
amounts to having a mixed state between the intended bosonic state 
and other bosonic states.  
Such a noise
does not exhibit a strong dependence 
on the computational basis but rather it depends 
on the intrinsic properties of the simulated state.

The feasibility of quantum computers is closely related to 
efficient versions of the Church-Turing thesis. This thesis originated in 
the works of Church and Turing in the mid-1930s is now commonly 
formulated as follows:

\begin {itemize}
\item[]
{\bf The Church-Turing Thesis:} 
{\em A  Turing machine can simulate any realistic model of 
computation.}
\end {itemize}

Another formulation of the Church-Turing thesis (following Kamal Jain) 
which does not rely on the 
notion of a Turing machine is simply: {\em Every realistic model of 
computation can be simulated on your laptop.}
The Church-Turing thesis is closely related to a fundamental distinction 
between computable and uncomputable functions, a distinction which 
is the basis 
of the theory of computability. Another major distinction, between 
tasks that can {\it efficiently} be solved by computers and tasks that 
cannot, is the basis of computational complexity theory. The most 
famous conjecture in this theory asserts that 
non-deterministic polynomial time computation defines a 
strictly larger class of 
decision problems than polynomial-time 
computation, or, briefly,  $NP \ne P$.   
Itamar Pitowsky was one of the pioneers to study the connections between the 
Church-Turing thesis and physics, and to consider efficient versions 
of the thesis. The various possible versions of the efficient 
Church-Turing thesis touch 
on several fundamental problems regarding the physical origin of probability, 
classic and quantum. 
We will discuss the Church-Turing thesis and its efficient versions 
in Section \ref {CTH}.


Our  second direction is described in Section 
\ref {s:dn}. 
We propose and discuss two postulates on the nature of errors in
highly correlated noisy physical stochastic systems. The first postulate 
asserts that errors for a pair of  
substantially correlated elements
are themselves substantially correlated. The second postulate asserts that 
in a noisy system with many highly correlated elements there will 
be a strong effect of error synchronization. 
The basic idea is that
noisy highly correlated data cannot be stored or manipulated. 
On a heuristic level this conjecture is interesting for both the quantum 
and the classical cases.\footnote {Note that in
the classical case correlations do not increase the computational
power. When we run a randomized computer program, the random bits can 
be sampled once they are created, and it is of no computational 
advantage in the classical case to ``physically maintain'' highly correlated 
data.} In order to put these conjectures on formal grounds 
we found it necessary to restrict them to the quantum case and refer to 
decoherence, namely the information loss of quantum systems.
We indicate how to put our conjectures regarding entanglement and 
correlated noise on formal mathematical grounds. Details can be 
found in \cite {K,K:wna}.

In Section \ref {s:np} 
we discuss 
quantum evolutions that allow noise to accumulate. This is our 
third direction.
Here again we assume no structure on 
the Hilbert space of states.
The class of noisy quantum processes we describe is an 
interesting {\it subclass}
of the class of all noisy quantum processes described by time-dependent 
Lindblad equations.
The model is based on the standard time-dependent Lindblad equation 
with a certain additional ``smoothing'' in time. 
In Section \ref {s:ph} we discuss some  
physical aspects of our conjectures. 
In Section \ref {s:ita} I describe several places where Itamar Pitowsky's 
academic journey interlaced with mine going back to our days as students in the 1970s.



\section {Quantum computers, noise, fault tolerance, 
and the threshold theorem}
\label {s:bm}

\subsection {Quantum computers}

This section provides background on the models of 
quantum computers and noisy quantum computers. 
We assume the standard model of quantum computer 
based on qubits and gates with pure-state evolution.
The state of 
a single qubit $q$ is described by a unit vector $u = a|0>+b|1>$  in 
a two-dimensional complex space ${\cal H}_q$. 
(The symbols $|0>$ and $|1>$ can be thought of as 
representing two elements of a basis in $U_q$.) 
We can think of the qubit $q$ as representing 
$`0`$ with probability $|a|^2$ and $`1`$ with probability $|b|^2$. 
The state of a quantum computer with $n$ qubits 
is a unit vector in a complex Hilbert space $\cal H$: the $2^n$-dimensional 
tensor product of two-dimensional complex vector spaces for the 
individual qubits. 
The state of the computer thus 
represents a probability distribution on the $2^n$ strings
of length $n$ of zeros and ones. 
The evolution of the quantum 
computer is via ``gates.'' Each gate $g$ 
operates  
on $k$ qubits, and we can assume $k \le 2$. 
Every such gate represents a 
unitary operator on the ($2^k$-dimensional) tensor product 
of the spaces that correspond to these $k$ qubits. At every ``cycle time''
a large number of gates acting on disjoint sets of qubits operates.
We will assume
that measurement of qubits that amount to a sampling of 0-1 strings 
according to the distribution that these qubits represent is the final step 
of the computation.

\subsection {Noisy quantum computers}

The basic locality conditions for noisy quantum computers 
assert that the way in which the 
state of the computer changes between 
computer steps is approximately 
statistically independent for different qubits. 
We will refer to such changes as ``storage errors'' or ``qubit errors.'' 
In addition, the gates that carry the computation itself are imperfect.
We can suppose that every such gate involves a small number 
of qubits and that the gate's
imperfection 
can take an arbitrary form, 
and hence the errors (referred to as ``gate errors'') created 
on the few qubits 
involved in a gate can be statistically dependent. 
We will denote as ``fresh errors'' the storage errors 
and gate errors in one computer cycle.
Of course, qubit errors and gate errors propagate along the computation.
The ``overall error'' describing the gap 
between the intended state of the computer and its noisy state   
takes into account also the cumulated effect of errors from 
earlier computer cycles.

The basic picture we have of a noisy computer is that 
at any time during the computation 
we can approximate 
the state of each qubit only up to some small error term 
$\epsilon$. 
Nevertheless, under the assumptions concerning the errors 
mentioned above, computation is possible. The noisy physical qubits
allow the introduction of logical ``protected'' qubits that are 
essentially noiseless.
We will consider the same model of quantum computers 
with more general notions of errors. We will study more general 
models for the fresh errors. 
(We will not distinguish between gate errors and storage errors, the two 
different components of fresh errors.)
Our models 
require that the storage errors 
should not be statistically independent (on the contrary, they should 
be very dependent) 
or that the gate errors should not be 
restricted to the qubits involved in the gates and that they should be 
of sufficiently general form. 


\subsection {The threshold theorem}

We will not specify the noise at each computer cycle but rather 
consider a large set, referred to as the {\it noise envelope}, of quantum 
operations the noise can be selected from. 

Let $\cal D$ be the following envelope 
of noise operations for the fresh errors: 
the envelope for storage errors ${\cal D}_s$ will consist of 
quantum operations that have a tensor product 
structure over the individual qubits.
The envelope for gate errors ${\cal D}_g$ will consist of quantum 
operations that have a tensor product 
structure over all the gates 
involved in a single computer cycle (more precisely, 
over the Hilbert spaces representing the qubits in the gates). 
For a specific gate the noise can be an arbitrary quantum operation on 
the space representing the 
qubits involved in the gate. (The threshold theorem concerns 
a specific universal set of gates $\cal G$ that
is different in different versions of the theorem.)

\begin {theo} [Threshold theorem] \cite{AB2,Kit1,KLZ}
Consider quantum circuits with a universal set of gates $\cal G$. 
A noisy quantum circuit with a set of gates $\cal G$ and noise envelopes 
${\cal D}_s$ and ${\cal D}_g$ is capable of effectively 
simulating an arbitrary noiseless quantum circuit, 
provided that the error rate for every 
computer cycle is below a certain threshold $\eta>0$. 
\end {theo} 

The threshold theorem, which was proved by three  
independent groups 
of researchers,
is a major scientific achievement. 
The proof relies  on the notion of quantum error correcting 
codes and some important constructions for such codes, and it requires
further deep and difficult ideas.  
The value of the threshold in original proofs of the threshold theorem was 
around $\eta =10^{-6}$ and it has since been improved by at least one 
order of magnitude.  
Numerical simulations based on new fault tolerance schemes (\cite{Kn}) 
suggest that the value of $\eta$ can be raised to 3\%.

\section {Codes}
\label {s:codes}

A quantum code is a very fundamental object.  
A code is simply a subspace $\cal L$ of the Hilbert space $\cal H$ 
describing 
the states of a quantum system. States in $\cal L$ are called ``codewords.'' 
Note that this description does not refer 
to any additional structure of the underlying 
Hilbert space but in order to talk about errors and 
error-correction we need to restrict our attention to 
Hilbert spaces that have tensor product structure. For that we first 
talk about classical error-correcting code.

The construction of (classical) error-correcting codes 
is among the most celebrated applications of mathematics. 
Error-correcting codes are eminent in today's technology from 
satellite communications to computer memories.
A binary code $C$
is simply a set of 0-1 vectors of length $n$.
Recall that the Hamming distance between two such vectors $x$ and $y$ 
is the number of coordinates $x$ 
and $y$ differs. 
The minimal 
distance $d(C)$ of a code $C$  is the minimal 
Hamming distance between 
two distinct elements $x,y \in C$. 
The same definition applies when 
the set $\{0,1\}$ is replaced by a larger alphabet $\Sigma$. When 
the minimal distance is $d$ the code $C$ is capable 
of correcting  $[d/2]$ arbitrary errors. 




Consider the $2^n$-dimensional Hilbert space $\cal H$ describing 
the states of a quantum computer with $n$ qubits. The tensor product structure
of $\cal H$ enables us to talk about error-correction capabilities of the 
quantum code $\cal L$. The first known example of a quantum 
error-correcting code 
by Shor \cite {S2} and Steane \cite {St} was on seven qubits. 
It was capable to correct arbitrary single-qubit errors.

The hope regarding FTQC is that no matter what the quantum computer 
computes or simulates, 
nearly all of the noise will be a mixture of states that are 
not codewords in the error-correcting code, but which are 
correctable to states in the code.  In contrast we made the 
following conjecture: 
 
\begin {quote}
{\bf Conjecture 1:} {\em The process for creating a quantum error correcting 
code will necessarily lead to a mixture of the 
desired codeword with undesired codewords.} 
\end {quote}

Note that this conjecture does not rely on a tensor product structure 
of states of a quantum computer and is quite general. It is especially 
appealing if the process for creating the quantum code does not involve 
fault tolerance. The motivation behind this conjecture is simple. 
The process for creating 
an error-correcting code can be seen as implementing a function 
$f:(x_1,x_2,\dots,x_k) \to (y_1,y_2,\dots,y_n)$. (Typically, $n$ is  
larger than $k$.) 
The function $f$ encodes $k$ qubits (or $k$ bits in the classical 
case) using $n$ qubits.
A noisy vector of inputs will lead to a mixture of the target codewords 
with undesired codewords.

Here is an example: {\it Kitaev's toric code} is a remarkable 
quantum error-correcting code \cite {Kit1}
 described by a vector space $\cal L$ 
of states of qubits placed on the vertices of  
an $n$ by $n$  lattice on a two-dimensional torus. The toric code 
encodes a pair of qubits.
Kitaev's toric codes can be regarded as a quantum analog of a 
one-dimensional
Ising model.
If we use a toric code for quantum computation, the hope 
supported by the standard noise model, is that during the 
computation we will have a mixed state that can be correctable 
(except for an exponentially small probability) to a unique codeword. 
Conjecture 1 implies that we can only achieve 
mixed states that can be correctable 
to a (non-atomic) probability distribution of codewords. 
A way to describe these different views is to associate to a noisy toric state
seven real parameters. One parameter represents the distance to the 
Hilbert space of toric states\footnote{in terms of the expected 
number of qubit errors}, and six parameters represent the 
encoded two qubits. A noisy state in the conventional picture will amount to 
having the first parameter described by a certain probability distribution 
and the other six parameters by a delta function! 
Quantum fault tolerance will enable
the creation of such states. In contrast, 
Conjecture 1 asserts that a noisy toric code state will be described by 
a probability distribution 
supported on a full seven-dimensional set in terms of these seven parameters.

Classical error-correcting codes are {\em not} special cases of quantum codes 
and Conjecture 1 does not apply to them. Nevertheless it is  
interesting to ask to what extent Conjecture 1 is consistent with classical
computation. 
A convenient and fairly realistic 
way to think about classical computation is by 
regarding the computer bits as representing a low-temperature Ising 
model on a large number of particles. Each particle has an 
up spin or a down spin and the interactions
force the entire system to be in one of two states. 
In one of these states every particle is, with a high probability, 
an up spin, and in 
the other state every particle is, with a high probability, a down spin.
One way to think about the Ising model is as follows. The Ising 
state represents the probability 
for an up spin. This is a sort of classical analog to a quantum code: 
For every real parameter $p$ 
the ``codewords'' representing $p$ are configurations where a 
fraction of $p$ particles have up spins. 
The classical computer is built from Ising-model-based bits. Each bit thus
represents a probability distribution. The gates allow us to ``measure''
this probability distribution and create according to the values 
for one or two bits new such bits representing the basic Boolean operation.  
Note that in this probabilistic description we have storage and gate noise
that is compatible with (a classical version of) Conjecture 1. Nevertheless, 
this allows noiseless classical computation.


\section {The Church-Turing Thesis and efficient computation}
\label {CTH}
 
The Church-Turing thesis asserting that ``everything computable is 
computable by a Turing machine,'' and its sharper 
forms about efficient computation, can be regarded as laws of physics. 
However, there is no strong connections between the thesis and 
computability in general and theoretical physics. 
When it comes to the original form of the Church-Turing thesis 
(namely when efficiency of computation is not an issue), it does 
not seem to matter if you allow quantum mechanics or work just 
within classical mechanics. However, for a model of 
computation based on physics it is important to specify 
what are the available approximations or, in other words, the 
ways in which errors are modeled. (Failure to do so may 
lead even to ``devices'' 
capable of solving undecidable problems.) There are various 
proposed derivation of the Church-Turing thesis from physics laws. 
On the other hand, there are various ``hypothetical physical worlds'' 
that are in some tension with the Church-Turing thesis (but 
whether they contradict it is by itself an interesting 
philosophical question). 
A paper by Pitowsky \cite {Pi} deals with 
such hypothetical physical worlds. See also \cite {PS}.

{\it Efficient computation} refers to a computation 
that requires a polynomial number of steps
in terms of the size of the input. 
The efficient Church-Turing 
thesis, first stated, as far as I know, by Wolfram \cite {Wo} in the 80s, 
reads:


\begin {itemize}
\item[]
{\bf Classical Efficient Church-Turing  Thesis:} 
{\em A  Turing machine can efficiently simulate 
any realistic model of computation.}

\end {itemize}

One of the most important developments in computational 
complexity in the last four decades 
is the introduction of randomized algorithms. Randomized algorithms 
use some internal randomization and it is quite surprising
that this allows for better algorithms for various 
computational tasks. This leads to

\begin {itemize}
\item[]
{\bf Probabilistic Efficient Church-Turing Thesis:} 
{\em A probabilistic Turing machine can efficiently simulate 
any realistic model of computation.}
\end {itemize}


The postulate of quantum computation is in conflict with these versions
of the efficient Church-Turing thesis. 
The analogous conjecture 
for quantum computers is 

\begin {itemize}
\item[]
{\bf Quantum Efficient Church-Turing Thesis:} {\em 
A quantum Turing machine can 
efficiently simulate any realistic model of computation.}
\end {itemize}

One aspect of the efficient Church-Turing thesis 
(again, both in its classical, probabilistic, and quantum versions) 
is that it appears to imply that NP-complete problems cannot 
be computed efficiently 
by any computational device. This again is a physics conjecture of a 
sort (and it depends, of course, on mathematical conjectures from 
computational complexity). 
There are some interesting attempts to relate  
the impossibility of solving NP-complete problems to physics, but still 
the interaction with theoretical physics is rather slight. 
Another interesting aspect of (both classic and quantum versions of) 
the efficient Church-Turing thesis is the implication 
that physical models that require infeasible 
computations are unrealistic. It is an interesting question 
in the context of 
model-building. 
Should we take computational complexity into consideration
in scientific modeling? Can a model that is not efficiently 
computable, or not computable at all, still be useful?

What then is the correct description of the computational 
power supported by our physical reality? When it comes to randomization 
there is strong support to the assertion 
that weak sources of randomness can replace perfect random sources. 
There is some support to the stronger assertion that randomization is not 
really required and can be replaced by deterministic 
processes called {\em pseudorandom generators}. Of course, 
even if randomness does not 
genuinely enhance the computational power, 
which is a common belief among experts,  
randomized algorithms yield important heuristic 
ways to design powerful algorithms. 
Indeed, the use of randomness is 
quite substantial
even in algorithmic processes in nature.  
When it comes to quantum algorithms, there is 
strong supporting evidence to the 
assertion that quantum 
computers are computationally superior compared to classical ones. 
Therefore, the feasibility of quantum 
computers is strongly related to the correct model of 
computation for our physical world.
The threshold theorem asserts that noisy quantum computers
where the noise level is sufficiently low, and the noise satisfies
some natural assumptions, 
have the full computational power 
of quantum computers (without noise). 

We end this section with a question relating the interpretation 
of probability to computation. 
One of the approaches to the foundations 
of probability, classical and quantum alike, 
regards the world as deterministic and probability as expressing 
only human uncertainty. 
We can ask, 
if any probability in nature (be it classic or quantum) 
only expresses human uncertainty, how can using probability
genuinely enhance the computational power? 


\section {Noisy stochastic correlated physical systems}
\label {s:dn} 
\subsection {The postulate of noisy correlated pairs}
\label {s:p1}

The purpose of this section is to propose and discuss the
following postulate:

\begin {itemize}
\item [{\bf [P1]}]
{\em Any noisy physical system, 
is subject to noise in which the errors for a pair of elements
that are substantially statistically dependent
are
themselves substantially
statistically dependent.}
\end {itemize}

Another way to put Postulate [P1] is: noisy correlated elements
cannot be approximated up to almost independent error terms: if
we cannot have an approximation better than a certain error rate
for each of two correlated elements, then an uncorrelated or almost
uncorrelated approximation is likewise impossible.



We now formulate a related conjecture for quantum computers:

\begin {quote}
{\bf Conjecture 2:} {\em A noisy quantum computer is subject to noise 
in which information leaks for two substantially
entangled
qubits have a substantial positive 
correlation.}
\end {quote}


{\bf Remarks:}

\noindent
1. {\bf Real-life examples: The weather 
and the stock market. }
We can discuss Postulate P1 for cases of (classical) stochastic
systems with highly correlated elements.  I am not aware of a case of
a natural system with stochastic highly correlated elements that
admits an approximation up to an ``almost independent'' error term.
This is the kind of approximation required for fault-tolerant
quantum computation. Can we expect to estimate the 
distribution of prices of two very
correlated stocks in the stock market up to an error distribution
that is almost independent? Or take, for example, the weather. 
Suppose you wish to forecast
the probabilities for rain in twenty nearby locations. We
suppose these probabilities will be strongly dependent. Can we
expect to have a forecast that is off by a substantial error
that is almost statistically independent for the different locations? 

To make this question a little more formal, consider not how accurately
a weather forecast  
predicts the weather,  
but rather how it  
predicts  (or differs from) a later weather forecast.
Let $\cal D$ be the distribution that represents the best forecast we
can give for the rain probabilities at time $T$ from the data we
have at time $T-1$. Let $\cal D'$ be the best forecast from
data we have at time $T-1-t$. 
Suppose that $\cal D$ is highly
correlated. Postulate [P1] asserts that we cannot 
expect that the difference ${\cal D} - {\cal
D'}$ will be almost statistically independent for two locations
where $\cal D$ itself is substantially correlated.

\noindent
2. {\bf The threshold theorem and pair purification.} The
threshold theorem that allows FTQC has various remarkable
applications, but Conjecture 2 can be regarded as challenging its
simplest consequence. The assumptions of 
the threshold theorem allow the errors on a pair of qubits 
involved in a gate to be statistically dependent. In
other words, the outcome of a gate acting on a pair of qubits 
prescribes the position of the
two qubits only up to an error that is allowed to exhibit an
arbitrary form of correlation. The  
process of fault tolerance
allows us to reach pairs of entangled qubits that, while still being
noisy, have errors that are almost 
independent.
Note that fault tolerance does not 
improve the ``quality'' of individual qubits, and fault-tolerant computation 
allows computation in noisy computers where at any point 
the state of an individual qubit can only be 
estimated up to a certain small error.

\noindent
3. { \bf Causality.} We do not propose that the entanglement of the pair of noisy qubits 
{\it causes} the dependence between their errors. The correlation between
errors can be caused, for example, by the process leading to the entanglement
between the qubits, or simply by the ability of the device to achieve
strong forms of correlation.  

\subsection {The postulate of error synchronization}
\label {s:es}

Suppose we have an error rate of $\epsilon$. The assumptions of
the various threshold theorems (and other proposed methods for 
quantum fault tolerance) imply that the probability of a
proportion of $\delta$ qubits being ``hit'' is exponentially small
(in the number of bits/qubits)
when $\delta$ exceeds $\epsilon$. Error synchronization refers to an
opposite scenario: there will be a substantial probability of a
large fraction of qubits being hit. 


\begin {itemize}
\item [{\bf [P2]}]
{\em In any noisy physical system with many substantially correlated
elements there will be a strong effect of spontaneous error-synchronization.}
\end {itemize}

For noisy quantum computers we conjecture:

\begin {quote}
{\bf Conjecture 3:} {\em In any quantum computer 
at a highly entangled state there will be a
strong effect of 
error-synchronization.}
\end {quote}


{\bf Remarks:}

\noindent
1. {\bf Empiric.} Conjectures 2 and 3 can be tested 
for quantum computers with a small number of qubits
(15-30)\footnote{``Why not just test Conjecture 2 on two qubits?'', a reader 
may ask. 
The reason is that for gated pairs of entangled qubits 
the assertion of Conjecture 2 
is consistent with standard noise models. Using fault tolerance to create 
entangled qubits with uncorrelated noise requires a larger number of qubits.}
Even if such devices where the qubits themselves are
sufficiently stable are still well down the road, they are  
to be expected long before the superior complexity power of quantum computers 
kicks in. 


\noindent
2.  {\bf Spontaneous synchronization for highly correlated systems.} 
Spontaneous synchronization of physical systems is a well known phenomenon. 
See Figure 1 demonstrating spontaneous synchronization 
of metronomes\footnote{The picture 
is taken from http://www.youtube.com/watch?v=DD7YDyF6dUk\&feature=related}. The
idea that for the evolution of highly correlated systems changes
tend to be synchronized, so that we may witness rapid
changes affecting large portions of the system (between long
periods of relative calm), is appealing and may be related to other 
matters like sharp
threshold phenomena and phase transition, the theory of evolution, 
the evolution of
scientific thought, and so on. Spontaneous synchronization 
is also related to the issue of pattern
formation for correlated systems.

The idea that errors or troubles tend to synchronize is also familiar. 
This idea is conveyed in the Hebrew proverb ``When troubles come 
they come together.''\footnote{There are similar proverbs in various other 
languages, see: 
http://gilkalai.wordpress.com/2010/03/06/when-it-rains-it-pours.} 
We can examine the possibility of error
synchronization for the examples considered above.
Can we expect synchronized errors for weather forecasts? Can we
expect stock prices, even in short time scales, to exhibit
substantial probabilities for changes affecting a large proportion
of stocks? 

\begin{figure}
\centering
\includegraphics[scale=0.7]{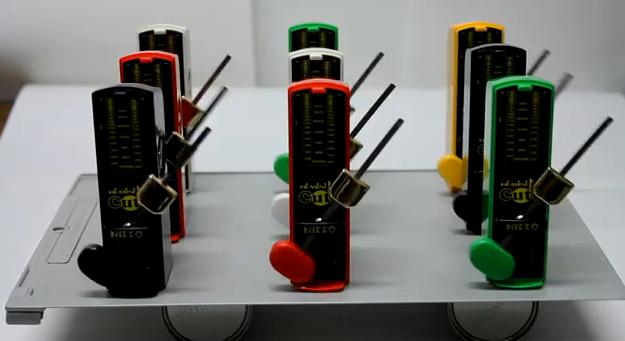}
\caption{spontaneous synchronization of metronomes}
\label{fig:0}
\end{figure}

\noindent
3. {\bf Error synchronization and the concentration of measure
phenomenon.} A mathematical reason to find spontaneous
synchronization of errors an appealing possibility is that  it
is what a ``random'' random noise looks like. Talking about a random
form of noise is easier in the quantum context. If you prescribe
the noise rate and consider the noise as a random (say unitary)
operator (conditioning on the given noise rate), you will see a perfect
form of synchronization for the errors, and this property will be
violated with extremely low probability.\footnote{ Random unitary operators
with a given noise rate are {\it not} a realistic form of noise. 
However, the fact that perfect error-synchronization 
is the ``generic'' form of noise may suggest that
stochastic processes describing the noise will approach this
``generic'' behavior unless they have  good reason not to.}



\noindent
4. {\bf Probability, secrets, and computing.}
We will now describe a 
difficulty for our postulates at
least in the classical case. Consider a situation where Alice
wants to describe to Bob a complicated correlated distribution $\cal
D$ on $n$ bits that can be described by a polynomial-size
randomized circuit. Having a noiseless (classical) computation
with perfect independent coins, Alice can create a situation 
where for Bob the distribution of the $n$ bits is 
described precisely by $\cal D$. 
In this case the values of
the $n$ bits will be deterministic and $\cal D$ reflects Bob's
uncertainty. 
Alice can also make sure that for Bob the distribution of the $n$
bits  will be ${\cal D} + {\cal E}$, where $\cal E$ describes independent
errors of a prescribed rate.


Is this a counterexample to our Postulates [P1] and [P2]? One can
argue that the actual state of the $n$ bits is deterministic and
the distribution represents Bob's uncertainty rather than  ``genuine''
stochastic behavior of a physical device.\footnote {Compare the
interesting debate between Goldreich and Aaronson \cite {GA} on
whether nature can ``really'' manipulate exponentially long
vectors.} 
But the meaning of ``genuine stochastic behavior of a
physical device'' is vague and perhaps ill-posed. 
Indeed, what is the
difference between Alice's secrets and nature's secrets? In any case, 
the difficulty described in this paragraph cannot be 
easily dismissed.\footnote {The distinction
between the two basic interpretations of probability as either expressing
human uncertainty or as expressing some genuine physical
phenomenon is an important issue in the foundation of (classical)
probability. 
Opinions range from
not seeing any distinction at all between these concepts to
regarding human uncertainty as the only genuine interpretation.} 
The formulation of Conjectures 2 and 3 is especially 
tailored to avoid this difficulty.

\subsection *{The mathematical forms of the Conjectures 2 and 3}

For a formal mathematical description of Conjectures 2 and 3 the reader 
is referred to \cite {K,K:wna}.
The best way we found for expressing formally
correlation between information leaks (Conjecture 2) and 
error synchronization (Conjecture 3) is 
via the expansion of quantum operations 
representing the noise in terms of multi-Pauli operations. 
The basic measure of entanglement for a 
pair of qubits in joint pure state is 
in terms of the von Neumann entropy. 
It is useful to consider {\it emergent entanglement} which is the maximum 
expected entanglement for two qubits after
we separably measure (and look at the outcome of) the 
remaining qubits of the computer. 
We can define the notion of highly entangled state 
in terms of emergent 
entanglement, and if we strengthen Conjecture 2 to 
deal with a pair of qubits 
with high emergent entanglement then we can prove that this 
already implies Conjecture 3; see \cite {K}. It is an interesting question
to identify quantum codes for which Conjecture 1 implies error synchronization.

\section {When noise accumulates}
\label {s:np}

A main property of FTQC is that it enables us to suppress noise 
propagation: the effect of the noise at a certain computer cycle 
diminishes almost 
completely already after a constant number of computer cycles.  
In this section we 
would like to formally model quantum systems for which noise 
propagation is not suppressed. For more details and discussion 
see \cite {K:wna}.

A way to force un-suppressed noise propagation into the model 
is as follows.
Start with an ideal unitary quantum 
evolution $\rho_t: 0 \le t \le 1$ on some Hilbert space $\cal H$. Suppose 
that $U_{s,t}$ denotes the unitary operator describing 
the transformation from time $s$ to time $t$, ($s<t$). 
We will use the same notation $U_{s,t}$
to denote the  
quantum operation
extending the action of $U_{s,t}$ to mixed states. 
$\rho_t$ is thus described by the abstract Schrodinger equation
\begin {equation}
\label {Schrodinger}
d \rho / dt = -i[H_t,\rho].  
\end {equation}

\noindent
Next consider a noisy version where $E_t$ is a 
superoperator   
describing the infinitesimal noise at time $t$. 
This data 
allows us to describe the noisy evolution $\sigma_t$ via the time-dependent-
Lindblad equation\footnote{Note that time-dependent Lindblad 
equations as defined here form a very 
general class of evolutions. In the literature, 
Lindblad evolutions often refer only
to the time-independent (Markovian) case or to other restricted classes.} 

\begin {equation}
\label {Lindblad}
d \sigma / dt = -i[H_t,\sigma] + E_t(\sigma).  
\end {equation}

\begin{figure}
\centering
\includegraphics[scale=0.75]{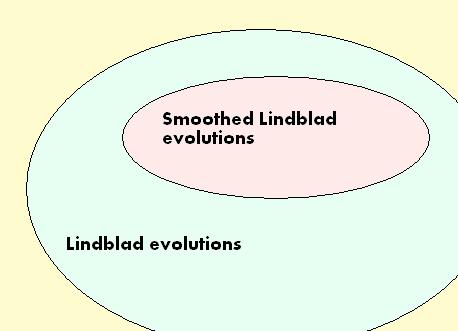}
\caption{Smoothed (time-dependent) Lindblad evolutions are a restricted 
subclass of the class of 
all (time-dependent) Lindblad evolutions}
\label{fig:2}
\end{figure}

\noindent
We will now describe a certain ``smoothing'' in time of the noise. 
Let $K$ be a positive continuous function on [-1,1]. 
We write 
$\bar K(t) = \int_{t-1}^tK(s)ds$. 
Replace the noise superoperator $E_t$ at time $t$ by 
\begin {equation}
\label {e:master}
\tilde E_t = 
(1/\bar K(t)) \cdot \int_0^1 K(t-s)  U_{s,t} E_s U^{-1}_{s,t}ds.
\end {equation}

\noindent
We denote by $\tilde \sigma_t$ the noisy evolution described by the smoothed 
noise superoperator
\begin {equation}
\label {smoothedLindblad}
d \tilde \sigma / dt = -i[H_t,\tilde \sigma] + \tilde E_t(\tilde \sigma).  
\end {equation}
We refer to such evolutions as {\it smoothed time-dependent Lindblad evolutions}.

We will restrict 
the class of 
noise superoperators  and 
we will suppose that $E_t$ and hence $E'_t$ are 
described by POVM-measurements (see \cite {NC}, Chapter 2).  

\medskip

{\bf Definition:} {\em Detrimental noise}
refers to noise (described by a POVM-measurement) 
that can be described by 
equation 
(\ref {e:master}).

\begin {quote}
{\bf Conjecture 4:} 
{\em Noisy quantum processes are subject to detrimental noise.}
\end {quote}
















\section {Physics}
\label {s:ph}





We now turn our attention to some physical aspects of the conjectures. 
One place to examine some suggestions of this paper is  
current implementations 
of ion-trap computers. In these implementations we need to move 
qubits together in order to gate them, and this may suggest that, in each 
computer cycle, an additional noise 
where errors are 
correlated for {\it all} pairs of qubits is in 
place.\footnote{For example, noise of periodic nature may get synchronized 
in a similar way to synchronized metronomes.} 
This is an example where properties of accumulated noise 
may occur for other reasons. 

Let us go back to the example of simulating bosonic states 
with a noisy quantum computer. Here the code is the Hilbert space 
of bosonic states and Conjecture 1 asserts that 
part of the noise 
is a mix of the intended bosonic state with 
other unintended bosonic states. 
This is different from local noise which depend on the computational bases. 
Noise accumulation seems consistent with the 
familiar property of physical systems
where the low-scale structure is not witnessed when we look at larger scales. 
We do not yet have quantum computers that simulate bosonic states but 
we do have several natural and experimental 
processes that come close to this description, like 
phonons, which can be regarded 
as a bosonic state (on a macroscopic scale) ``simulated'' on 
microscopic ``qudits.''\footnote{A qudit is like a qubit which is based on a 
Hilbert space of dimension not restricted to two.} 
Another relevant example is that of Bose-Einstein condensation on cold atoms. 
Describing the bosonic state in terms of individual atoms is 
analogous to describing a complicated state of a quantum computer 
in terms of the computational basis. This analogy enables us to ask 
if the deviation of a state created experimentally 
from a pure state can be described by independent noise 
operators on the different atoms. Conjecture 1 proposes 
a different picture, namely,
that a state created experimentally can be described as a
mixture of different pure Bose-Einstein states.
%
These examples 
can serve as a good place to examine noise.

\begin{figure}
\centering
\includegraphics[scale=0.8]{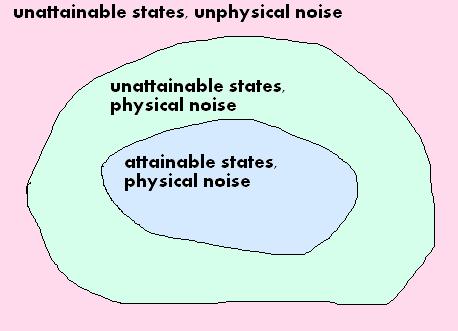}
\caption{Given a proposed architecture for a quantum 
computer it is possible that for some hypothetical states 
that cannot be achieved the proposed properties of noise are ``unphysical.'' 
The place to examine the conjectures is for attainable states.}
\label{fig:3}
\end{figure}

Several people have commented that  
our suggested properties of noise for some (hypothetical) 
quantum computer architecture 
at some quantum state $\rho$ allow instantaneous signaling, and thus 
violate basic physical principles. This is perfectly correct, but our
proposed conclusion 
is that this quantum computer architecture simply 
does not accommodate the quantum state $\rho$. (See Figure 3.)



Finally, an important proposal for implementing quantum computation is via topological quantum computing. 
Topological quantum computing is based on a 
remarkable class of quantum codes that represent 
certain representations of the braid group. The traditional bosons and fermions
can be replaced by objects called 
{\it anyons}.\footnote {Bosons are named after Bose and fermions 
after Fermi. Anyons are named after the 
word ``anything.''}
The extreme stability to noise expected 
for anyonic systems 
relies on similar 
assumptions to those enabling
quantum error correction. 
Conjecture 1 has a direct bearing on anyons, 
and the case for Conjecture 1 is especially strong for anyons because 
the proposed experimental processes for creating 
them do not exhibit quantum fault tolerance.
Our conjecture asserts that when we experimentally create anyons 
we will witness a mixture of the 
intended state with other states of the same type and we will not witness
the strong stability of certain anyonic systems that is predicted by 
current models. 


\section {Itamar}
\label {s:ita}

This paper is devoted to the memory of Itamar Pitowsky. 
Itamar was a great guy; he was great in science and great in the humanities. 
He could think and work like a mathematician, and like a physicist, 
and like a philosopher, and like a philosopher of science, 
and probably in various additional ways. And he enjoyed 
the academic business greatly, and took it seriously, with humor. 
Itamar had an immense human 
wisdom and a modest, level-headed way of expressing it. 
Itamar's scientific way and mine interlaced 
in many ways, a few of which are related to this paper. Itamar's 
approach to the foundation of quantum mechanics 
was that
quantum mechanics is a theory of non-commutative 
probability which (like classical probability) 
can be seen as a mathematical language for the other laws of 
physics. (This paper, to a large extent, adopts this point of view. 
But, frankly, I do not quite understand the other points of view.) 

In the late 70s when Itamar and I were both graduate 
students I remember him 
telling me enthusiastically about his thoughts on the Erd\H{o}s-Turan problem.
This is a mathematical conjecture that asserts 
that if we have a sequence of integers
$0<a_1<a_2<\cdots< a_n<\dots$ such that the 
infinite series $\sum \frac {1}{a_n}$ diverges then we can 
find among the elements of the sequence
an arithmetic progression of length $k$, for every $k$. (It is 
still unknown even for $k=3$.)  
Over the years both of us spent some time on this conjecture 
without much to show for it. 
Some years later
both Itamar and I got interested, for entirely different reasons, 
in remarkable geometric objects called cut polytopes. 
Cut polytopes are obtained by 
taking the convex hull of characteristic vectors of 
edges in all cuts of a graph. Cut polytopes 
arise naturally when you try to understand 
correlations in probability theory and Itamar wrote several seminal papers 
in their study; see \cite {P:cp}. Cut polytopes came to play, in an 
unexpected way, in the work 
of Jeff Kahn and myself where we disproved Borsuk's conjecture \cite {KK}.
Over the years, Itamar and I  got interested in 
Arrow's impossibility theorem \cite {Ar}
which Itamar regarded 
as a major 20th-century intellectual achievement. He gave a course centered 
around this theorem and years later so did I.
We were both members of the Center for the 
Study of Rationality at the Hebrew University and 
we both participated 
in a recent interesting debate regarding 
installing a camera in the Center's kitchenette  which turned out to 
raise interesting questions regarding privacy, shaming, and cleanliness, 
\cite {eum}. 

The role of skepticism in science and how (and if) 
skepticism should be practiced
is a fascinating issue. It is, of course, 
related to this paper which describes skepticism 
over quantum fault tolerance, widely believed to 
be possible. (I also believe it might be possible,
but I think we should explore how it can be impossible.)
Itamar and I  had long 
discussions about skepticism in science, 
and about the nature and practice of scientific debates.
A few years ago Itamar gave a lecture 
about physicists' feeling after the standard model was found 
that a complete understanding of the 
fundamental laws of physics is around the corner. This feeling was manifested, 
according to Itamar, in Weinberg's wonderful 
book {\em The First Three Minutes: A Modern View of the 
Origin of the Universe}. 
Itamar described some later developments in physics 
that, to some extent, shattered this euphoric feeling. 
Later, following his visit to the Perimeter Institute and a 
conversation he had with Lee Smolin, 
Itamar told me about the skeptical
blogs and books about string theory. 
I got interested and eventually wrote an 
adventure book \cite {K:g} 
criticizing 
the skeptical approach of the ``string war'' skeptics. 
Much earlier, in the early 90s, we had several conversation 
regarding the ``Bible code'' research.
Three researchers wrote a paper that showed statistical evidence
for hidden codes in the Bible. Neither Itamar nor I believed this claim for a minute
but it certainly led to interesting issues regarding statistics, philosophy of science, scientific ethics, and more.
Itamar was convinced that science can cope with such claims and saw no harm in them being published.
Some years later, I was part of a team that offered \cite {MBBK} statistical biased 
selection as a much simpler and more familiar explanation for 
the outcomes of this research.


A  week before Itamar passed away, Itamar,  Oron Shagrir, and I sat at our little CS cafeteria and talked about probability. Where does probability come from?
What does probability mean? Does it just represent human uncertainty? Is it just an emerging mathematical concept that 
is convenient for modeling? Do matters 
change when we move from classical to quantum mechanics? When we move to quantum physics the 
notion of probability itself changes for sure, but is there a change in the interpretation of 
what probability is?  A few people passed by and listened, and it felt like this was a direct continuation of 
conversations we had had while 
we (Itamar and I; Oron is much younger) were students in the early 70s. 
This was to be our last meeting.

\end {document}